\documentclass{appolb}
\usepackage{graphicx}
\usepackage{hyperref}

\begin{document}
\title{Exclusive four-pion photoproduction in ultra-peripheral heavy-ion collisions at RHIC and LHC energies%
\thanks{Presented at XXVI Cracow EPIPHANY Conference, LHC Physics: Standard Model and Beyond}%
}
\author{Mariola K{\l}usek-Gawenda$^{(1)}$, J. Daniel Tapia Takaki$^{(2)}$
\address{$^{(1)}$Institute of Nuclear Physics Polish Academy of Sciences, PL-31342 Krakow, Poland}
\address{$^{(2)}$Department of Physics and Astronomy, The University of Kansas, Lawrence, KS, USA}
}
\maketitle
\begin{abstract}
We study the photoproduction of exclusive $2\pi^+2\pi^-$ mesons in ultra-peripheral heavy-ion collisions at RHIC and LHC energies. Predictions in photon-nucleus interactions are 
calculated for various resonances at central and forward rapidities. The recent H1 
preliminary data are utilized to improve the description of the poorly known 
$\gamma p \to 4\pi^\pm p$ process. We present the comparisons of 
our results to the available STAR data at RHIC, and made predictions for LHC energies.  
\end{abstract}
  
\section{Introduction}
There has been a renewed interest in the study of exclusive production of four charged pions. 
The $\gamma p \to 2\pi^+2\pi^- p$ reaction is an interesting process from the point of view of resonance production, including searchers for exotic resonances as well as searches for non-linear QCD gluon saturation effects. Recently the H1 collaboration at HERA has presented preliminary data~\cite{H1_DIS2018} on exclusive four-pion production. This represents an opportunity to better understand the underlying physics and the various possible resonances contributing to this decay. Exclusive production of $4\pi^\pm$ at the HERA collider was studied at small photon virtualities ($Q^2<2$~GeV$^2$), \textit{i.e.} photoproduction. The first observation of $\pi^+\pi^-\pi^+\pi^-$ photoproduction 
in ultra-peripheral heavy-ion collisions was reported by the STAR Collaboration~\cite{Abelev:2009aa}. This state was observed at the low transverse momentum, \textit{i.e.} coherent photoproduction. The invariant mass spectrum shows a rather broad peak at $M_{4\pi} = 1540 \pm 40$~MeV with a width of $\Gamma_{\rho \to 4\pi} = 570 \pm 60$~MeV. These values characterize $\rho^0(1700)$ vector meson. Motivated by the preliminary data recently presented by the H1 collaboration, we have studied four charged pion photoproduction in ultra-peripheral heavy-ion collisions with the aim to make predictions for LHC energies.  

\section{Photonuclear process}
A calculation of vector meson photoproduction 
can be obtained using the vector meson dominance model
\cite{Sakurai:1960ju,GellMann:1961tg}. We consider a simple version of the vector dominance model (VDM) where the photon fluctuates into hadronic component that then interacts with the proton via the Pomeron or Reggeon exchange. The proton, meson or photon cross sections ($X=p, V, \gamma$ respectively) have been successfully described using the Donnachie and Landshoff model~\cite{Donnachie:1992ny}. 
The dependence of the total cross section with two main trajectories
takes the following expression 
\begin{equation}
\sigma_{tot}(Xp) = 
\alpha_1 W_{\gamma p}^{-\delta_1} +
\alpha_2 W_{\gamma p}^{\delta_2} \,.
\label{eq:sig_PomReg}
\end{equation}

The component with a negative power corresponds to the Reggeon exchange, more precisely, $\rho, \omega, f$ and $a$ exchange). The second term in Eq.~\ref{eq:sig_PomReg} arises from
the Pomeron pole exchange. At sufficiently high energies, only the Pomeron trajectory
is important. The cross section for exclusive vector meson production 
can be determined by analogy to the Reggeon/Pomeron exchange, 
and a fit to $\sigma(\gamma p \to X p)$ takes the form equivalent to Eq.~\ref{eq:sig_PomReg}. Recently the CMS Collaboration has measured exclusive $\rho^0(770)$ meson photoproduction in ultra-peripheral p--Pb collisions at $5.02$~TeV centre-of-mass energy~\cite{Sirunyan:2019nog}, reporting the measured $\delta_1$ and $\delta_2$ which are used in this analysis.
\begin{figure}[!h]	
\centerline{
		\includegraphics*[scale=0.4]{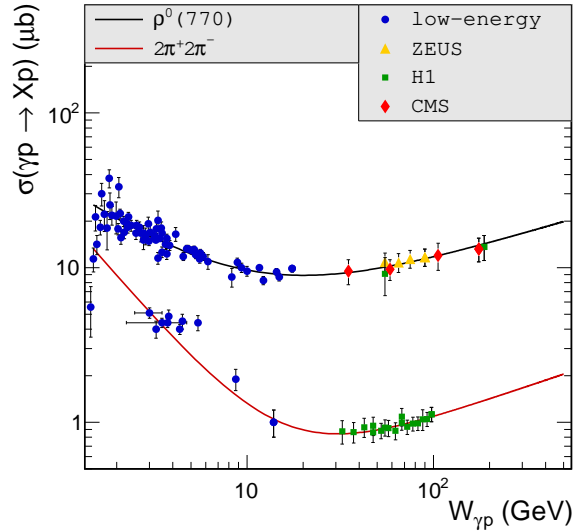}}
	\caption{Total cross section for the exclusive production of
		$\rho^0(770)$ vector meson (upper line) and $2\pi^+2\pi^-$ production (lower line).
		The parametrization of low-energy 
		\cite{Cassel:1981sx, Adams:1997bh, Hadjidakis:2004zm, Morrow:2008ek} 
		and high-energy $\rho^0(770)$ data \cite{Aid:1996bs,Breitweg:1997ed,Sirunyan:2019nog}
		as well as low-energy \cite{Bingham:1972az,Davier:1973fy,Schacht:1974fq,Atiya:1979ip,Aston:1981zz} 
		and preliminary H1 four-charged-pion data \cite{H1_DIS2018} are presented.}
	\label{fig:sig_gp_Vp}
\end{figure}

Fig.~\ref{fig:sig_gp_Vp} shows the total cross section for exclusive photoproduction
of $\rho^0(770)$ vector meson and the $2\pi^+2\pi^-$ state. 
The low-energy data (blue circular points), $W_{\gamma p}<20$~GeV,  
are obtained using fixed-target experiments.
The cross section for $\sigma(\gamma p \to \rho^0(770)p)$ process 
was measured at $<W_{\gamma p }> = 92.6$~GeV by the CMS Collaboration and found to be 11.2 $\pm$ 1.4 (stat) $\pm$ 1.0 (syst.)~$\mu$b. 
The CMS measurement is compatible with the previous data~\cite{Cassel:1981sx, Adams:1997bh, Hadjidakis:2004zm}, while performing the energy dependence of the momentum-transfer distribution for the firs time. There is a good agreement between data and our calculation, especially for HERA and LHC energies. 

Because of the lack of experimental data on four-pion production there has not been a consensus on which of the possible resonance states plays the most important role.
At small $W_{\gamma p}$ energies ($<3$~GeV), the non-resonant signal dominates, while excited states of $\rho$ meson start to become important for $W_{\gamma p}>4$~GeV. Such resonances ought to be considered in the context of $\pi^+\pi^-\pi^+\pi^-$ production at HERA, RHIC and LHC energies. 
A simple fit to H1 data~\cite{H1_DIS2018} shows that the sum of the Breit-Wigner $\rho(1600)$ resonant
\footnote{At present, the $\rho(1570)$ signature is used 
	\cite{Tanabashi:2018oca} instead of $\rho(1600)$.}, 
non-resonant four-pion state and complex phase 
($\rho(1570)-4\pi$ interference) describes fairly well the experimental points.
While from the four-pion invariant mass distribution, $M_{4\pi}$, one can conclude that a single broad $\rho(1570)$ resonance is the most significant
resonance in the photoproduction of exclusive $2\pi^+2\pi^-$ final states, a correlation in the four-pion invariant mass of oppositely charged pions is observed in data. There seems to be an enhancement of the four-pion signal around $M_{4\pi} = 1450$ and $1700$~MeV. This is defined as the $\rho(1450) \equiv \rho^\prime$ and 
$\rho(1700) \equiv \rho^{\prime\prime}$ states. Unfortunately, these resonances do not have their $4\pi$ branching rate determined to the desired precision. Thus, we can only try to estimate this factor. The branching ratio for $\rho \to e^+e^-$ is another value that is not well determined at present. Table~\ref{tab:res} shows the mass and width of resonances \cite{Tanabashi:2018oca} as well as the estimated values of $\Gamma( V \to e^+e^-)$ that are used in this analysis.

\begin{table}[!h]
	\caption{Characteristic of the $\rho^{\prime}$, $\rho(1570)$ and $\rho^{\prime \prime}$ mesons.}
	\label{tab:res}
	\centerline{
	\begin{tabular}{|c|c|l|l|} \hline 
		Resonance		& m [GeV]	& $\Gamma$ [GeV]	& $\Gamma_{e^+e^-}$ [keV]	\\ \hline 
		$\rho(1450)$	& 1.465		& 0.4				& 4.30 - 10 					\\ 	
		$\rho(1570)$	& 1.570		& 0.144				& 0.35 - 0.5					\\ 
		$\rho(1700)$	& 1.720		& 0.25				& 6.30 - 8.9					\\ \hline
	\end{tabular} }
\end{table}

In the VDM-Regge approach the photon state $|\gamma>$ is characterized as a quantum mechanical superposition
of the quantum electrodynamics photon state $|\gamma_{QED}>$ and a hadronic state $|h>$:
\begin{equation}
|\gamma> = \mathcal{N}|\gamma_{QED}> + |h> \mbox{, where } |h> = \sum\limits_{h} \frac{e}{f_V}|V>
\label{eq:gamma}
\end{equation}
with $f_V^2 = \frac{e^2 \alpha_{em} m_V }{3\Gamma(V\to e^+e^-)}$ 
being the vector meson -- photon coupling. 
This constant factor is expressed through the $e^+e^-$ decay width of the vector meson
with the $m_V$ mass. In addition, $f_V$ does not depend on the $Q^2$ 
and gives a probability for the transition of the photon to the vector meson.
The $\mathcal{N}$ in Eq.~(\ref{eq:gamma}) is a normalization factor.
The hadronic state $|h>$ is assumed to have the same additive quantum number as the photon.
In the case of vector meson, this condition means that $J^{PC} = 1^{--},$ and $Q=B=S=0$.
The bare QED component cannot interact with hadrons \cite{Bauer:1977iq}.
It is also worth noting that the most significant contributions 
of the hadronic component of $|V>$
come from the light vector mesons ($\rho^0$, $\omega$ and $\phi$) 
and this thesis constitutes the main hypothesis of the VDM approach.
The simple VDM-Regge approach allows to describe the transition between photon
and vector meson using the following relation of the photon-proton cross section:

\begin{eqnarray}
\sigma_{tot}(Vp) &=& \frac{f_V^2}{e^2} \sigma(\gamma p \to V p) 
\label{eq:sig_Vp}
\end{eqnarray}

Eq.~(\ref{eq:sig_Vp}) is appropriate only for the limit virtuality $Q^2<m_V^2$. 
In the VMD-Regge model, the $Vp$ total cross section 
can be calculated using the optical theorem
taking into account the forward $\gamma p \to V p$ cross section:
\begin{eqnarray}
\sigma_{tot} (Vp) =  \frac{f_V}{e} \sqrt{\frac{16 \pi}{1+\eta^2} \frac{\mathrm{d} \sigma (\gamma p \to V p; t=0)}{\mathrm{d} t}},  
\end{eqnarray}
where $\eta$ is the ratio of the real to the imaginary part of 
scattering amplitude. 
Since H1 does not report the differential cross sections in the momentum
transfer square for the $\gamma p \to 2\pi^+ 2\pi^- p$ process, we have taken 
$\sigma(\gamma p \to 2\pi^+ 2\pi^- p)$ as a starting point in this analysis. 
It turns out our simple calculation can describe the experimental data on $\sigma(\gamma p \to 2\pi^+ 2\pi^- p)$
(see Fig.~\ref{fig:sig_gp_Vp}), especially at H1 energies.

Photoproduction of vector mesons on the nucleus 
can be considered by combining the VDM-Regge model
and the Glauber theory of multiple scattering \cite{Glauber_book}.
The total cross section for light vector meson---nucleus interaction is calculated using the following formula \cite{Klein:1999qj}:
\begin{eqnarray}
\sigma_{tot}(VA) = \int \left[ 1 - \exp \left( -\sigma_{tot}\left( Vp \right) T_{A}(\vec{b}) \right)   \right] \mathrm{d}^2b
\label{eq:sig_tot_VA_cm}
\end{eqnarray}
where
\begin{eqnarray}
T_{A}(\vec{b}) = \int_{-\infty}^{+\infty} \rho(\vec{b},z) \mathrm{d}z \;.
\label{eq:TA}
\end{eqnarray}
The two-parameter Fermi model is used to describe the Au and Pb density \cite{DeJager:1987qc}. 
The charge distribution in the nucleus is normalized to the mass number by the relation $\int \mathrm{d}^2b  \rho(b,z) \mathrm{d}z=A$. Eq.~(\ref{eq:sig_tot_VA_cm}) is written in the so-called ``classical
mechanics" framework. It is worth noting that also the ``quantum expression"
is often used in the literature. However, as discussed in Ref.~\cite{Klusek-Gawenda:2015hja}, 
for coherent $J/\psi$ production in ultra-peripheral collisions there is a difference about $15 \%$ between 
these two different rescattering treatments. Moreover, the variance between the two approaches becomes larger for $\rho^0(770)$ photoproduction~\cite{thesis}. 
The cross section of coherent vector meson photoproduction on nuclei reads:

\begin{eqnarray}
\sigma \left( \gamma A \to VA \right) = \frac{1}{16 \pi} \frac{e^2}{f_V^2} \sigma^2_{tot}\left( VA \right) \int \left| F(t) \right|^2 \mathrm{d}t \;.
\label{eq:gA_VA}
\end{eqnarray}
The nuclear form factor is expressed through charge distribution in the nucleus. 
Here we use the so-called realistic form factor. See  \textit{e.g.} Ref.~\cite{KlusekGawenda:2010kx} for more details.

\begin{figure}[!h]	
	\centerline{
	\includegraphics*[scale=0.34]{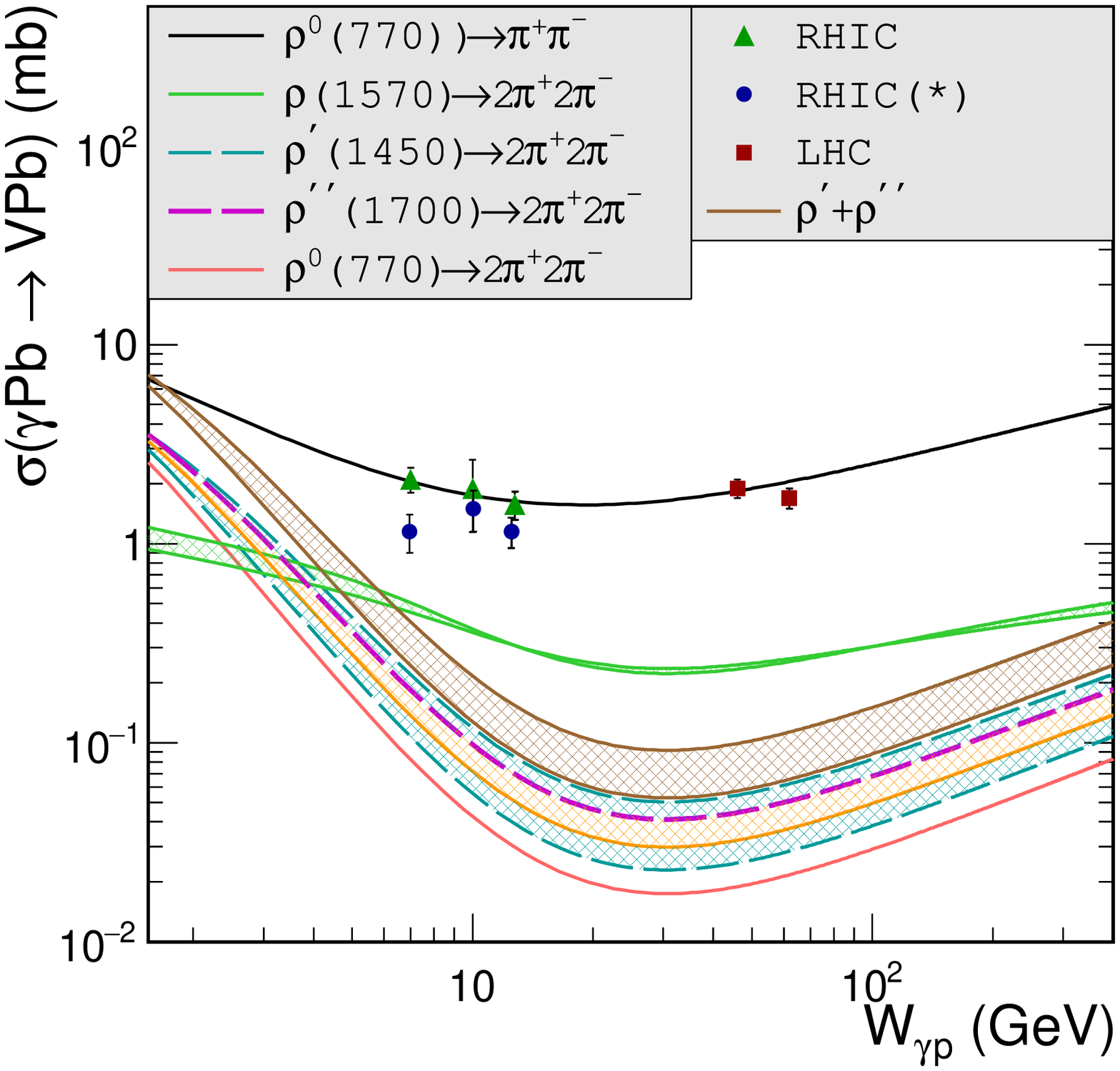}
	\includegraphics*[scale=0.34]{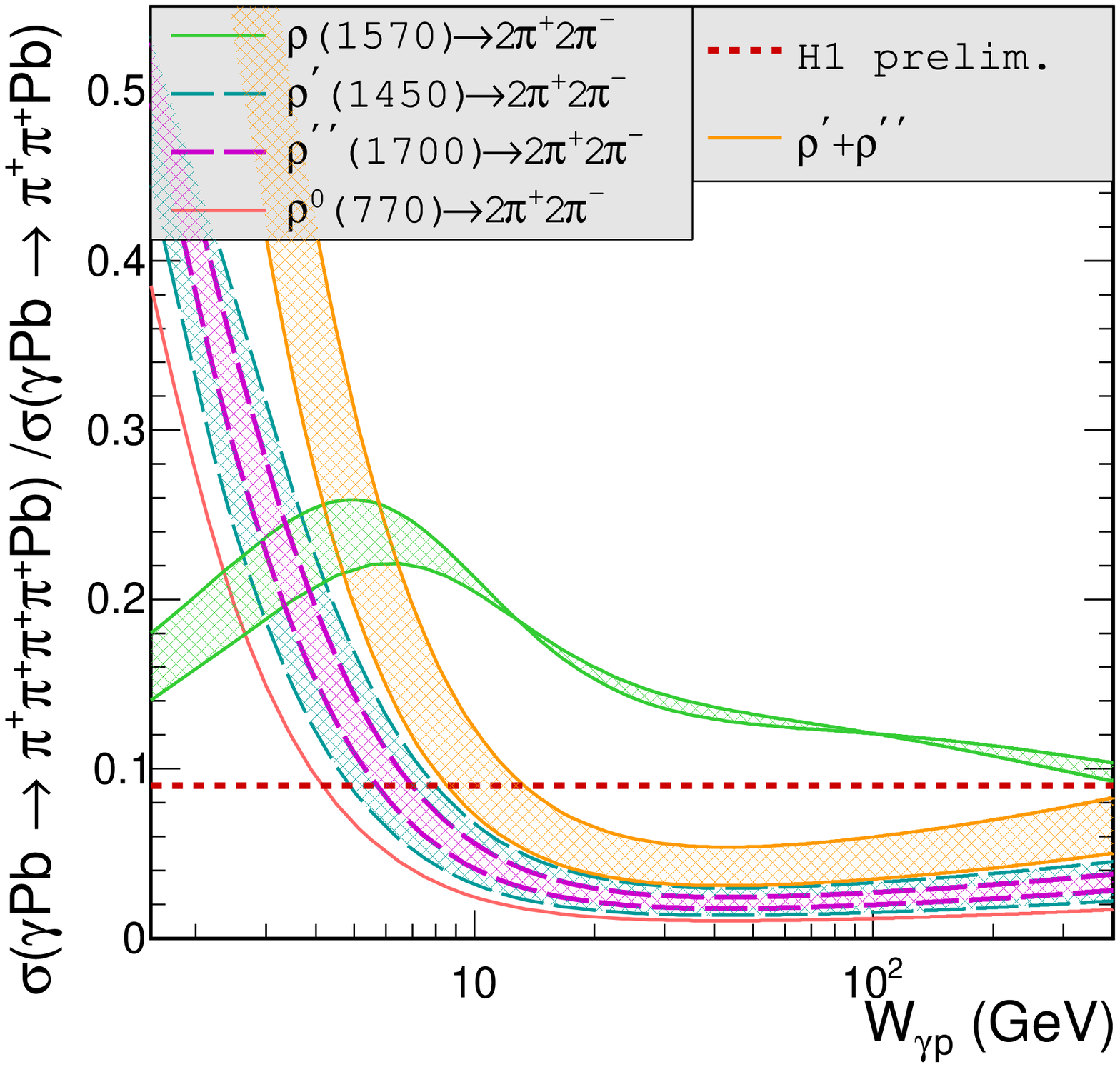}}
	\caption{Left panel: 
		Cross section for the $\gamma Pb \to V Pb$ process.
		The upper black line shows the results for $V=\rho^0(770)$
		that decays into two charged pions. Lowers curves
		correspond to $2 \pi^+ 2 \pi^-$ production 
		(more details in the main text).
		Results for $\gamma Pb \to \pi^+\pi^- Pb$
		are compared with existing experimental data
		\cite{Adler:2002sc,Abelev:2007nb,Agakishiev:2011me,Adam:2015gsa}.
		Right panel: Ratio of four to two charged pions photoproduction. The theoretical results are compared with the recent H1 preliminary data~\cite{H1_DIS2018}.}
	\label{fig:sig_gA_VA}
\end{figure}

The left-hand side of Fig. \ref{fig:sig_gA_VA} shows the cross section 
for photoproduction of two- and four-pion state.
Results are presented for the lead nucleus.
The largest cross section appears for the case when
$\rho^0(770)$ meson decays into $\pi^+\pi^-$ channel.
The theoretical result which includes the classical approach
of $\rho^0$ photoproduction (Eq.~(\ref{eq:sig_tot_VA_cm})) has been found to give a good description of the LHC experimental data. We present two sets of data by the STAR collaboration at RHIC. The first one (green triangular points) have been used to calculate $\rho^0(770)$ photoproduction in a number of theoretical/phenomenological studies, \textit{e.g.} Ref.~\cite{Frankfurt:2015cwa}. Our parametrization is consistent with such calculations, while Ref.~\cite{Cepila:2018zky} reports a small discrepancy with these approaches. A new points constitutes an enumeration which was obtained by estimation of STAR data for the gold-gold collision energy which is equal to $62.4$, $130$, $200$ GeV. The cross section at mid-rapidity reads:
\begin{equation}
\sigma(\gamma A \to VA; y=0) = \frac{1}{N_{\gamma A}(y=0)} \frac{\mathrm{d}\sigma(AA \to AA V; y=0)}{\mathrm{d}y} \;.
\end{equation}
A new data set is calculated using a more realistic form factor which is designated in the flux of photon.
The flux of equivalent photon strongly depends on the charge distribution in the nucleus, particularly at small impact parameters. A detailed treatment of this issue can be studied in a future work. We know turn to the discussion of Fig.~\ref{fig:sig_gA_VA}. We note that the branching ratio of excited $\rho^0(770)$ states are poorly known. The lower curves and ranges correspond to $\rho^{\prime}, \rho^{\prime\prime}, \rho(1570)  \to 4\pi^\pm$ decay. The limits are determined by varying the $\Gamma_{e^+e^-}$ width range. At low photon-nucleus energies the sum of the $\rho^{\prime}$ and $\rho^{\prime\prime}$ contribution overcomes over the $\rho(1570) \rightarrow 4\pi$ result. The $\gamma Pb \to \rho(1570) Pb$ cross section starts to dominate over other excited states only from $W_{\gamma p}>8$~GeV. The lowest curve corresponds to $\rho^0(770)$ meson which decay into four-charged-pion channel.
The difference between the distribution for two- and four-pion states from the $\rho^0$ decay can be as large as of two order of magnitude from each other. The right panel of Fig. \ref{fig:sig_gA_VA}
shows the ratio od $\pi^+\pi^-\pi^+\pi^-$ to $\pi^+\pi^-$ photoproduction.
Theoretical results are compared to the recent H1 preliminary data  which is equal to $9 \%$ \cite{H1_DIS2018}. It is worth noting that the range of $W_{\gamma p}$ energy
which is used to photoproduction at LHC energy, is about $10$~GeV for $y=-4$ and $650$~GeV for $y=4$. The energy comes to $90$~GeV for $\rho(1570) \to 2\pi^+2\pi^-$ at mid-rapidity.
Although we do not have an excellent agreement with H1 value,
we are close to this number. The result for $\rho(1570)$ tends to overestimate the data
and the $\rho^{\prime}$ + $\rho^{\prime\prime}$ sum is smaller than H1 preliminary point,
in the range of energy which corresponds to LHC measurements.   

\section{Cross sections for the AA case}

Nuclear photoproduction of a vector meson $V$ can be written as a convolution
of the photonuclear cross section Eq.~(\ref{eq:gA_VA}) and equivalent photons fluxes:
\begin{equation}
\frac{\mathrm{d}\sigma(AA \to AA V)}{\mathrm{d}^2b \mathrm{d}y} = 
\omega_1 N(\omega_1,b) \sigma(\gamma A_2 \to V A_2) + 
\omega_2 N(\omega_2,b) \sigma(\gamma A_1 \to V A_1) \;,
\label{eq:tot_cs}
\end{equation}
where $\omega_i = m_V/2 exp(\pm y)$ denotes energy of emitted photon, $b$ is impact parameter. We consider ultra-peripheral collisions that means that transverse distance between the center of nuclei is larger than the sum over radii of nuclei~\cite{Contreras:2015dqa}. The photon flux depends on the form factor, more precisely, on the charge distribution in the nucleus.  A detailed study of the model for the photoproduction of vector mesons can be found in
\cite{Klusek-Gawenda:2015hja,Klusek-Gawenda:2013dka}. Eq.~(\ref{eq:tot_cs}) allows 
to calculate total cross section and differential cross section as a function
of the rapidity of outgoing photon or impact parameter. The completes analysis should
include the kinematic of the decay product. We study four-pion production so we should 
take into consideration the rapidity of each outgoing pion. This is done by the inclusion
of the smearing of $\rho$ mesons. The widths of $\rho^\prime, \rho^{\prime\prime}$ and $\rho(1570)$ are known. The main part of the spectral shape of the vector meson is calculated from the Breit-Wigner formula:
\begin{equation}
\mathcal{A} = \mathcal{A}_{BW} \frac{\sqrt{m m_V \Gamma(m)}}{m^2 - m_V^2 +im_V\Gamma(m)} + \mathcal{A}_{bkg} \;.
\end{equation}
The mass-dependent width is parameterized as follows:
\begin{eqnarray}
\Gamma(m) = \Gamma_V \frac{m_V}{m} \left(\frac{m^2-4m_V^2}{m_V^2-4m_\pi^2}\right)^{\frac{3}{2}} \;.
\end{eqnarray} 
The $\mathcal{A}_{bkg}$ is interpreted as the $\pi^+\pi^-$ background. This factor describes
the enhancement of the left hand side of the resonance term and some smearing of right hand side term.
One would like to have data that include the rapidity of each decay particle to perform detailed kinematic study of the $1 \rightarrow 4$ process. Here we simply assume the $1 \rightarrow 4$ state as a $1\rightarrow 2\rightarrow 4$ process:
\begin{eqnarray}
&&\sigma(AA \to AAV \to AA\pi^+\pi^-\pi^+\pi^-, y_V) =  \mathcal{C} \times \\
&&\left[ \sigma(AA \to AAV \to AA\pi^+\pi^-\pi^+\pi^-;y_{\pi_1}y_{\pi_2}) \times \right. \\
&& \left.  \sigma(AA \to AAV \to AA\pi^+\pi^-\pi^+\pi^-;y_{\pi_3}y_{\pi_4}) \right] \;,\nonumber
\end{eqnarray}  
where $\mathcal{C}$ is the normalization constant which is different for each excited state of the $\rho$ meson.
This normalization constant is calculated as follows: 
\begin{eqnarray}
\mathcal{C} &=& 	2\pi\frac{|\mathcal{A}|^2}{ \int |\mathcal{A}|^2 dm_{V}} 
				\frac{\mathrm{d}\sigma(AA \to AA V \to AA \pi^+\pi^-\pi^+\pi^-)}{\mathrm{d}^2b\mathrm{d}y_{V}}
				\mathrm{d}m_{V} \mathrm{d}y_{V} \mathrm{d} b\nonumber \\
			&/& \left[\frac{\mathrm{d}\sigma(AA \to AA V \to AA \pi^+\pi^-\pi^+\pi^-;y_{\pi_1}y_{\pi_2})}{\mathrm{d}y_{V_1}\mathrm{d}m_{V_1}} (1-z_{\pi_1}^2) \mathrm{d}y_{V_1}\mathrm{d}m_{V_1}\right] \nonumber \\
			&/& \left[ \frac{\mathrm{d}\sigma(AA \to AA V \to AA \pi^+\pi^-\pi^+\pi^-;y_{\pi_3}y_{\pi_4})}{\mathrm{d}y_{V_2}\mathrm{d}m_{V_2}} 
				(1-z_{\pi_3}^2) \mathrm{d}y_{V_2}\mathrm{d}m_{V_2}\right] \;.
\label{eq:const}
\end{eqnarray}
A weight is introduced into the calculation: $\sin^2(\theta)$, where $\theta$ is the scattering angle
in the vector meson center-of-mass system ($z_\pi=\cos(\theta)$). We set the condition: $z_{\pi_1} = - z_{\pi_2}$ and $z_{\pi_3} = - z_{\pi_4}$.

\begin{figure}[!h]	
	\centerline{
	\includegraphics*[scale=0.34]{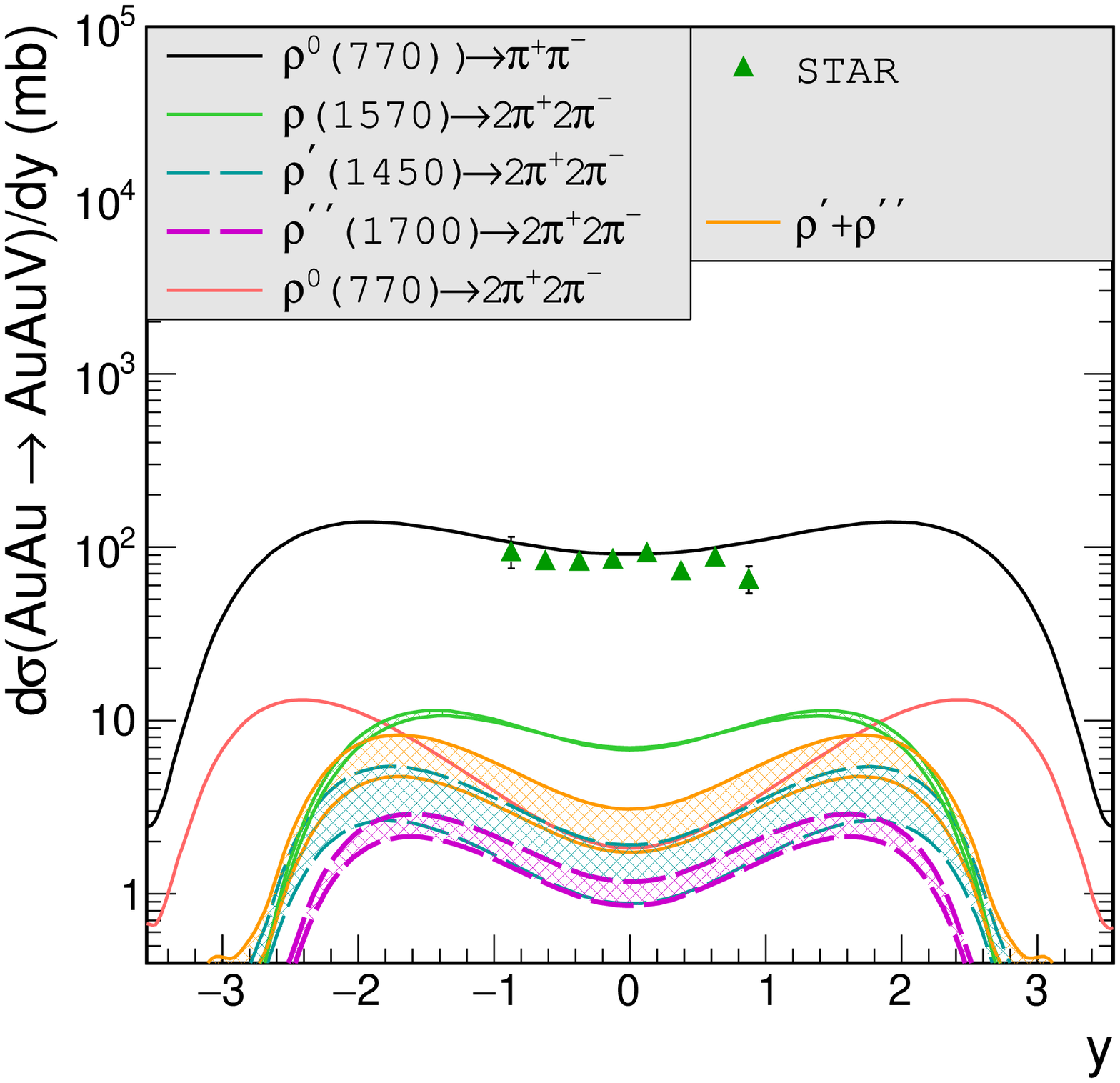}
	\includegraphics*[scale=0.34]{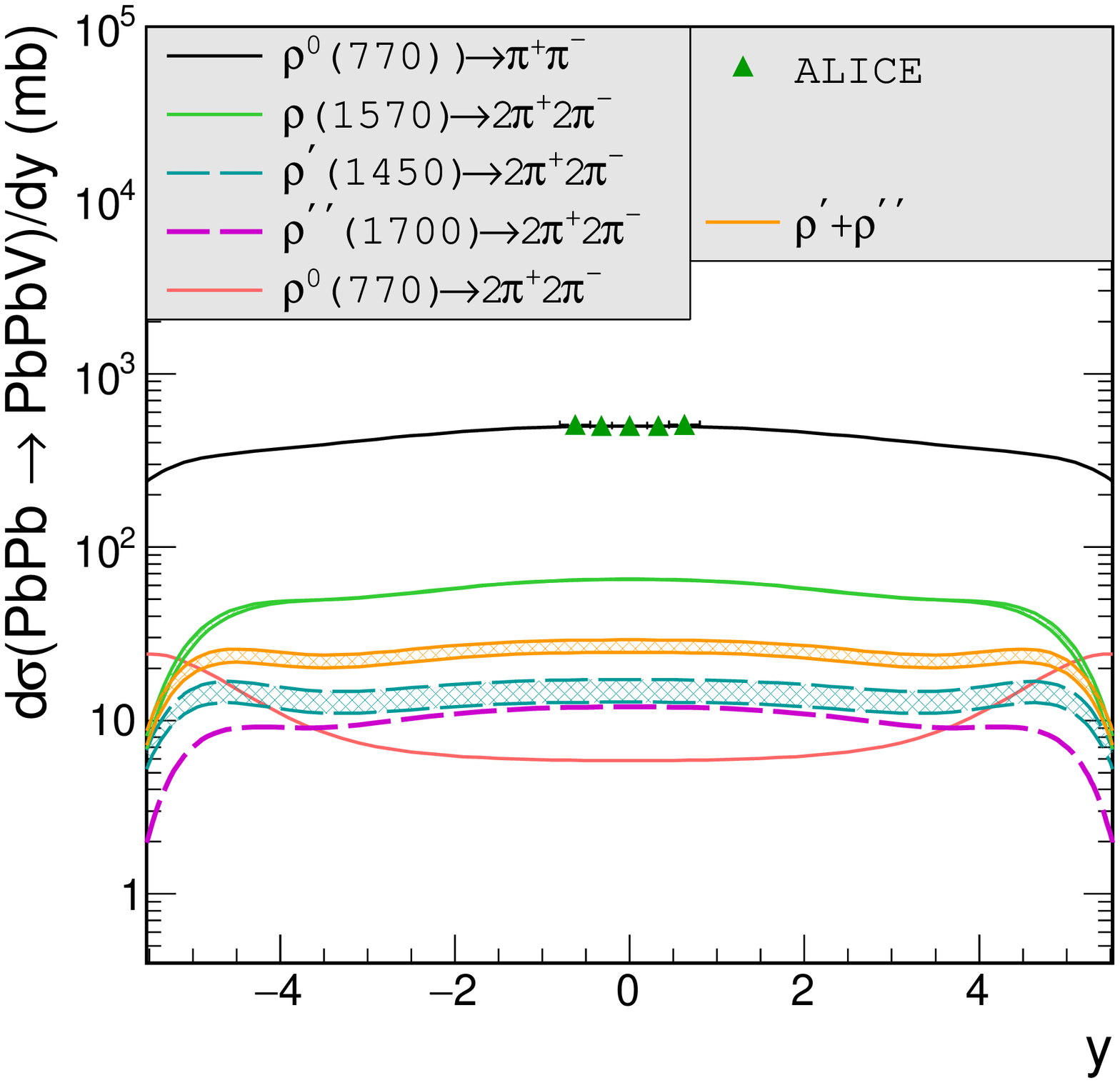}}
	\caption{Differential cross section as a function
		of rapidity of vector meson that decays into two or four
		charged pions. Results for $\pi^+\pi^-$ production are
		compared to the STAR data \cite{Abelev:2007nb} (left panel) and the ALICE data~\cite{Acharya:2020sbc} (right panel).}
	\label{fig:dsig_dy_V}
\end{figure}

Fig. \ref{fig:dsig_dy_V} presents results for 
ultra-peripheral Au--Au collisions at $\sqrt{s_{NN}}=200$~GeV (left panel)
and for Pb--Pb collisions at $\sqrt{s_{NN}}=5.02$~TeV.
Photoproduction of $\rho^(770)$ meson that decays into
two charged pions (the upper black line)
as well as photoproduction of vector meson 
($\rho(1570)$ - solid green line, $\rho(1450)$ - dashed blue line,
$\rho(1700)$ - dashed purple line, $\rho^0(770)$ - solid red line)
is considered. As done for the results presented in Fig.~\ref{fig:sig_gA_VA} the uncertainties associated to $V-\gamma$ coupling constant are shown. The contribution for the sum of the $\rho^{\prime}$ 
and $\rho^{\prime\prime}$ is also shown (orange band). One can observe that
the contribution for $\rho^0(770)\to \pi^+\pi^- \pi^+\pi^-$ state
plays an important role in the broad range of the meson rapidity at STAR and becomes less significant at forward rapidities.
The cross section for the production
of four-charged-pion is the largest in the case of
$\rho(1570)$ meson decay.

\begin{figure}[!h]	
	\centerline{
		\includegraphics*[scale=0.4]{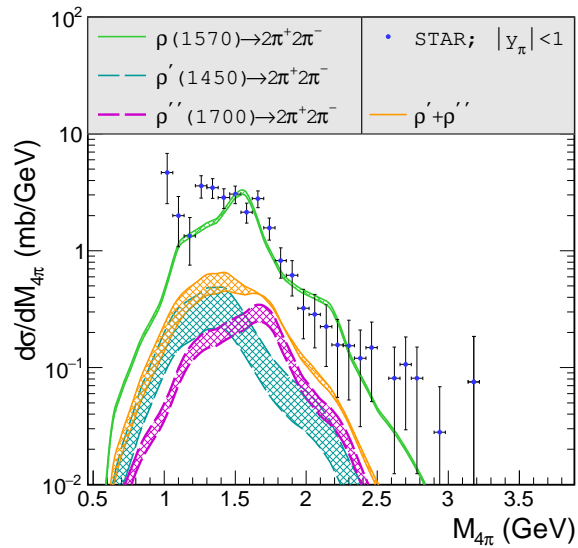}}
	\caption{Differential cross section as a function
		of four pion invariant mass. Data from the STAR collaboration are shown~\cite{Abelev:2009aa}.}
	\label{fig:dsig_dM_STAR}
\end{figure}

The STAR Collaboration has reported on the photoproduction
of four pions in ultra-peripheral Au--Au collisions at 
$\sqrt{s_{NN}}=200$~GeV. They have measured pions at mid-rapidity,
$|y_\pi|<1$. Fig. \ref{fig:dsig_dM_STAR} presents the four-pion invariant mass
distribution comparing the STAR data to our calculation. It also shows the contribution from the double-scattering $\rho^0$ mechanism discussed in Ref.~\cite{Klusek-Gawenda:2013dka}. Such a contribution
accounts for about $20 \%$ of the measured cross section. 
Four-pion production can be considered as a result of $\gamma\gamma \to \rho^0 \rho^0$
subprocess \cite{Klusek:2009yi}. However, the cross section for $Au Au \to Au Au \rho^0 \rho^0 \to Au Au 2\pi^+ 2\pi^-$
is about two orders of magnitude smaller than cross section measured by STAR.
Fig. \ref{fig:dsig_dM_STAR} also presents distributions for $\rho^\prime$, $\rho^{\prime \prime}$, the sum of these mesons and $\rho(1570)$. 
A correction for the acceptance function described in~\cite{Klusek-Gawenda:2013dka} is applied.
We observe a good agreement with the experimental data. In particular, the decay of $\rho(1570)$ resonance is found to give a good description of the STAR data. The shape of the four-pion invariant mass strongly depends
on the Breit-Wigner description. On other words the shape of the smearing of the resonance depends on the factor
which is responsible for the background correction. Regardless of the value of this factor the normalization of the smearing mass function is the same. While the H1 has also presented the four-pion invariant mass distribution, more data are needed to make direct comparisons between our calculation.  The sum of incoherent $\rho^\prime$ and $\rho^{\prime\prime}$ mesons gives the cross section in the range of $0.41-0.74$ mb. The coherent sum of the mesons includes two $\e^{i\varphi}$ factors in the Breit-Wigner formula.
The first one corresponds to the $\rho^\prime$ resonance and $\varphi_2$ is for $\rho^{\prime\prime}$. The shape of the invariant mass strongly depends on the phase. At the same time it appears that changing the parameters will not result in a cross section of almost one order of magnitude larger than that reported. The experimental cross section within $|y|<1$ is $\sigma = 2.4 \pm 0.2 \pm 0.8$~mb. The theoretical result for $Au Au \to Au Au \rho(1570)(\to \pi^+\pi^-\pi^+\pi^-)$ process gives a limit: $(2.16-2.31)$~mb. 

\begin{table}[!h]
	\caption{Total cross section for the production of exclusive $\pi^+\pi^-\pi^+\pi^-$ in ultra-peripheral Pb--Pb collisions at $\sqrt{s_{NN}}=5.5$~TeV.}
	\label{tab:tot_LHC}
	\centerline{
		\begin{tabular}{|c|c|l|l|} \hline 
			Resonance		& $|y_\pi|<1$	& $|y_\pi|<2.4$	& $2.5<y<4$ 	\\ \hline 
			$\rho(1450)$	& (1.1-2.3)		& (11-25)		& (0.2-0.5) 					\\ 	
			$\rho(1570)$	& (10-98)		& (103-105)		& (1.5-1.6)					\\ 
			$\rho(1700)$	& (1.6-2.2)		& (16-23)		& (0.3-0.4)					\\ \hline
Ref. \cite{Citron:2018lsq}	& 16			& 190			& 14					\\\hline
	\end{tabular} }
\end{table}
%

Since we obtain a good agreement between the STAR data and our calculation predictions to LHC energies are now discussed. Table~\ref{tab:tot_LHC} presents the total cross section for the photoproduction of exclusive four-charged-pion. 
Collision energies are for LHC energies. Results are presented for three intervals of the pion rapidity:
mid-rapidity in the range of (-1,1) and (-2.4,2.4), and at forward rapidity: (2.5-4). 
Our predictions are compared with those from the STARLight Monte Carlo generator~\cite{Citron:2018lsq}. 
A similar fraction for the two mid-rapidity intervals is obtained (within an order of magnitude), but the results that include the preliminary H1 total cross section for exclusive $2\pi^+2\pi^-$ production are smaller from those 
using the parametrization based on STAR data. 


\section{Summary and conclusions}

We have presented the results from a simple model for the photoproduction of vector mesons that decay into the pion states. By incorporating a parametrization of the recent preliminary data from the H1 Collaboration, this work permits to better understand the role of $\rho^\prime$ and $\rho^{\prime\prime}$ resonances as well as $\rho(1570)$ meson in exclusive four-pion production. Our study shows that the $\rho(1570)\to\pi^+\pi^-\pi^+\pi^-$ process dominates in the ultra-peripheral heavy-ion collisions. Only one $\rho^\prime$ or $\rho^{\prime\prime}$ resonances seem to be sufficient to describe the data. This work also shows the importance of the broad $\rho(1570)$ resonance into the $Au Au \to Au Au \pi^+\pi^-\pi^+\pi^-$ process. This work might serve future analyses of exclusive photoproduction of four-pion states at high energies.

\section*{Acknowledgements}
The study of MKG was partially supported by the Polish National Agency for Academic Exchange (NAWA) within the Bekker programme under grant PPN/BEK/2018/1/00424.
Research work of DTT is partly supported by the U.S. Department of Energy, Office of Nuclear Physics, Heavy Ion Nuclear Physics program, under Award DE-FG02-96ER40981.
MKG acknowledges useful discussion about four-body decay with Francesco Giacosa
during Epiphany2020 Conference.

\end{document}